# e-Installation: Synesthetic Documentation of Media Art via Telepresence Technologies

Jesús Muñoz Morcillo[a*], Florian Faion[b] [1], Antonio Zea[b] [2], Uwe D. Hanebeck[b] [3], Caroline Y. Robertson-von Trotha[a] [4]

[a] *ZAK | Centre for Cultural and General Studies, Karlsruhe Institute of Technology (KIT), Kaiserstr. 12, 76131 Karlsruhe, Germany*
[b] *Intelligent Sensor-Actuator-Systems Laboratory (ISAS), Karlsruhe Institute of Technology (KIT), Kaiserstr. 12, 76131 Karlsruhe, Germany*

**ABSTRACT**
In this paper, a new synesthetic documentation method that contributes to media art conservation is presented. This new method is called "e-Installation" in analogy to the idea of the "e-Book" as the electronic version of a real book. An e-Installation is a virtualized media artwork that reproduces all synesthesia, interaction, and meaning levels of the artwork. Advanced 3D modeling and telepresence technologies with a very high level of immersion allow the virtual re-enactment of works of media art that are no longer performable or rarely exhibited. The virtual re-enactment of a media artwork can be designed with a scalable level of complexity depending on whether it addresses professionals such as curators, art restorers, and art theorists or the general public. An e-Installation is independent from the artwork's physical location and can be accessed via head-mounted display or similar data goggles, computer browser, or even mobile devices. In combination with informational and preventive conservation measures, the e-Installation offers an intermediate and long-term solution to archive, disseminate, and pass down the milestones of media art history as a synesthetic documentation when the original work can no longer be repaired or exhibited in its full function.

1. ## Research Aims

The main aim of this research is to design a novel synesthetic documentation method for media artwork at risk under the perspective of "informational preservation" **[1].** For this purpose, advanced 3D modeling and telepresence technologies have been used, which allow a realistic immersive experience. This paper is a first step to improve conventional media art documentation not only for re-enactment purposes but also allowing permanent access to the virtualized artwork. In this way, the multimodal limitations of traditional audio-visual documentation methods such as video or photography are overcome. The goals within the involved research fields (media art conservation, advanced 3D modeling, and telepresence) include the enhancement of media art documentation on a synesthetic level and the development of improved techniques for immersive representation and interaction. In addition, the transversal effects of this work have influence on relevant research questions including the change of the authenticity concept in art conservation theory and the use of telepresence as an art creation tool.

---

* Corresponding author. Tel.: +4972160848933; fax: +4972160844811. E-mail address: jesus.morcillo@kit.edu
[1] florian.faion@kit.edu
[2] antonio.zea@kit.edu
[3] uwe.hanebeck@kit.edu
[4] caroline.robertson@kit.edu

## 2. Introduction

Media art has existed since the early 1960s. However, compared to traditional genres such as painting or sculpture, the lifespan of a piece of media art is very short: the technology it needs to operate is also the cause of its caducity. Moreover, museums are faced every day with the inexorable decline of technology-based artwork. Works of media art not only require constant maintenance but also take up much more exhibition space than museums can provide. As a result, they are often dismantled for maintenance and repair, or remain in the museum depot for long periods of time. When this happens, these pieces of art are no longer accessible to curators, art theorists, and the interested public. In this case, a detailed documentation that mostly consists of construction plans, interviews with the artists, and audio-visual material such as video or photography is the only way to ensure that these works of art can be examined. However, this kind of documentation cannot entirely reproduce the synesthetic experience level that media artwork such as video, sound art, kinetic sculptures, or media art installations requires to produce meanings. Curators and art theorists can only speculate on the full aesthetic impact of an artwork on this basis, unless that artwork is re-installed.

In the near future, art restorers will not be able to repair media artwork in accordance with satisfying authenticity criteria **[2]**. The reason for this is above all the obsolescence of technical components that are no longer being produced, such as CRT TVs and RGB projectors, CCF lamps or old data storage forms such as punched tapes or even old EPROMs. Given this scenario of a cultural heritage that is jeopardized and difficult to access, there is an urgent need for a new kind of documentation that allows, as much as possible, for a realistic representation of all the synesthesia levels implied in media artwork. Such documentation is necessary to protect and preserve the meanings and processes that might otherwise be lost along with the material work itself. Advanced 3D modeling and telepresence technologies can make a significant contribution in this regard.

As an anthropological category in the history of ideas, telepresence is a concept that can be traced all the way back to ancient times: the dream of an artificial life, an artistic tradition of virtual reality (e.g. life-size and immersive depictions), and the religious search for a disembodied conception of the human mind are the anthropological constants that converge in the idea of telepresence **[3]**. According to that, the human mind is natural predisposed toward immersive experience without simultaneously being incredulous of such experiences. Nevertheless, the definition of telepresence used in computer science research follows a less epistemological and much more technical notion as formulated by Sheridan (1989) **[4]**, who, assuming the human predisposition for telepresent experience, describes it as "the extension of a person's sensing and manipulation capability to a remote location." This "remote location" can also be a virtual world. According to both definitions, a carefully designed telepresence system would allow realistic access to and interactions with virtualized media artwork, in particular with those that are temporarily not available to the public, or those whose continuity cannot be guaranteed through current curatorial and conservation practices. The development of this new documentation method and its dissemination requires an interdisciplinary cooperation between experts in modern art preservation and documentation, experts in 3D modeling, telepresence technologies, and long-term archiving, as well as art communicators.

## 3. State of the Art

Since the end of the 1990s, there have been several international projects on the conservation and restoration of media art that bring the importance of documentation into focus as the first step to conserve and archive this new heritage. Several well-known projects and conferences about preservation and conservation of media art are "Modern Art: Who Cares?" **[5]**, "Seeing Double" **[6]**, "Inside Installations" **[7]**, the activities of the DOCAM Research Alliance **[8]**, as well as newer projects on the conservation of artwork created with computer technologies, like "Digital Art Conservation" **[9]**.

All these projects have one very important feature in common: they all regard media art documentation as an integral part of conservation strategies. From this point of view, it can be affirmed that documentation is also an indispensable part of the media art conservation process itself.

Institutions such as the Daniel Langlois Foundation and the INCCA Network have already performed pioneering work identifying conservation issues, observing artistic and curatorial practices, and proposing conservation strategies for the preservation of compromised art forms like media art installations, video sculptures, net art, and game artwork. The value of documentation for modern art conservation is also a commonplace in art restoration **[10]**. Good documentation requires well-founded knowledge about the piece of art in question focusing on conceptual and technological details and information about the intention of the artist and his or her expectations. In the year 2000, the art restorer Jon Ippolito published the "Variable Media Questionnaire" for ephemeral media art **[11]**. From today's perspective, it was the first attempt to involve media artists in conservation issues following a standard questionnaire similar to Erich-Ganzert Castrillo's detailed questionnaires and technical interviews with German painters **[12]**. Information about intention, future expectations of the artist, the optimal framework for exhibition, details about used technologies, and advice about how to preserve the artwork and what kind of replacements can be taken into account, help curators and art restorers, in their decision-making processes, to respect crucial authenticity criteria during conservation practice and reinstallation. Sometimes it is no longer possible to exhibit a media artwork in its original medium. In this case, it can be migrated or emulated. Ippolito distinguishes both strategies: "To migrate an artwork is not to imitate its appearance with a different medium, but to upgrade its medium to a contemporary standard, accepting any resulting changes in the look and feel of the work. To emulate an artwork, by contrast, is not to store digital files on disk or physical artifacts in a warehouse, but to create a facsimile of them in a totally different medium." **[11, p. 51]**

Making media art accessible to the public sometimes implies the need to vary some parameters of an artwork while still respecting the authenticity of its meaning. Migration and emulation are two documentation-based methods that allow the transmission of meaning at the expense of the original medium. In such cases, there is also often the need for a "reinterpretation" **[11, p. 52],** i.e., an adaptation of the art concept to the new medium. Both strategies can be included in the "informal preservation" model **[1]**, i.e., the preservation of meanings through documentation and migration, in opposition to the "preventive preservation" that tries to conserve all original parts of the artwork as long as possible with direct and environmental preservation measures. The "informational preservation" serves also as a frame for conservation strategies and as a starting point for the idea of virtualizing media artwork in order to create an e-Installation as a kind of "migrated" artwork.

Besides the inclusion of the artist in the conservation process, there are also descriptive methods that allow a personal perspective on the art experience like art depictions and video documentations. In combination with the technical and background information provided by the artist, it is possible to get a good idea of the whole artwork, although less intellectual effort would be needed and the findings would be more accurate with an intuitive experience of the artwork as is. Moreover, most video documentations with a conservation background tend to show a time-lapse recording of the set-up and dismantling of media art installations **[13],** while the available video documentations for the public do not even cover the full time length of media artwork.

On the one hand, we have to differentiate between the conservation and visualization of digital-born and virtual art, and digitization as a conservation mechanism. Single projects like "Aire ville Spatiale" **[14]** and the "Immaterial ArtStock Museum" **[15]** represent first attempts to collect and preserve digital-born 3D art in a digital space like Second Life, OpenSimulator, or realXtend. On the other hand, digitization has become a way to preserve and make accessible the content from old video art tapes **[16]** or to reconstruct archaeological finding places and reproduce historic buildings, pottery, or sculptures **[17, 18, 19, 20]**.

As for art visualization, the common opinion is that immersive virtual reality technologies (VR) offer very effective means to communicate cultural content, and are also effective for educational and presentation purposes **[21, 22]**. In the case of archaeology, the potential of 3D and augmented reality (AR) technologies for conservation issues have already been identified in the past **[18]**. These kinds of technologies, such as VR, AR, and Web3D, have, over the last ten years, mostly been used by science and archaeological museums that are interested in making their content attractive to the public **[23, 24]**. However, the so-called "virtual museums" are at best "content museums," i.e., websites with enhanced information in the form of pictures or videos. Genuine immersive platforms remain an exception.

Immersive hardware applications for cultural experience like the CINECA Virtual Theatre or the ReaCTOR of the Foundation of the Hellenic World also dedicate large exhibit spaces for their settings **[24]**. The ARCO platform (Augmented Representation of Cultural Objects) **[25]** uses interfaces to exploit multimodal visualization, but most of the VR devices being used in museums are desktop devices. Moreover, external devices like CAVEs (Cave Automatic Virtual Environment) or panoramic powerwalls are being used in modern museums to visualize new art forms or to complement the real museum's activities, but they are not being used as media art archives or for conservation or documentation purposes.

Nevertheless, there are already some VR systems that can interact with art in museums on the basis of commercial hardware such as "The Museum of Pure Form" (consisting of a CAVE and an exoskeleton with a haptic interface) or "The Virtual Museum of Sculpture" (panoramic powerwall). The disadvantages of these systems are that exoskeletons are heavy hardware and cannot easily be controlled by untrained operators, and that panoramic walls are very large and thus need large exhibit spaces. Since the use of head-mounted displays is not possible in combination with these systems, the participants cannot move about freely. Moreover, most of these projects (including exotic theater experiments with holographic illusions like "the Virtual Exploration of Turandot Stage" **[24]**) offer a non-interactive stereoscopic installation with movement and proprioceptive limitations for the participants.

For most media artwork – which is either at risk or rarely seen – there are documentation and conservation strategies in practice that do not consider the virtualization of the whole artwork as is but prefer a step-by-step preservation in order to keep the media artwork operating for as long as possible.

The virtualization of material parts integrating all digital software components and audio-visual signals, as well as all kinetic and interaction patterns, in a consistently playable, dynamic, and interactive 3D model would enable a new documentation method that allows telepresent accessibility to rarely exhibited or destroyed artwork to save the synesthetic level of experience and its structure of meanings. The benefits of synesthetic documentation for the conservation of the meaning and experience level of a media artwork was brought up by Muñoz Morcillo for the first time in 2011 in an essay on the documentation of changing media art **[26]**: "In this case study [Table Dancers by Stephan von Huene], one sees that the documentation of the change of a media artwork implies both a technical as well as a perception-related documentation. The interactive nature of the Table Dancers can mainly be found in the descriptions, no photo can document this fact. [...] Accessing this work would be virtually possible today if we had, e.g., an interactive 360° view of the installation and the ability to integrate its functions into a multimedia application [...]."

Our research continues and materializes this idea of the perception-related documentation, i.e., synesthetic documentation in the form of photogrammetrically comprehended, 3D-modelled and programmed artwork, and a suitable telepresence-based visualization of the virtualized media artwork using, e.g., head-mounted displays (HMD), body tracking systems, haptic interfaces, and "motion compression" algorithms, which are being developed **[27, 28, 29, 30]** at the Laboratory for Sensor Actuator Systems (ISAS) at the Karlsruhe Institute of Technology (KIT).

## 4. e-Installation: Telepresence as a Media Art Documentation Method

Multimodal devices and telepresence systems already allow a lifelike experience of virtual scenarios in a new kind of immersive virtual reality that implies genuine telepresence research topics like the plenoptic **[31]**, plenacoustic **[32],** and plenhaptic **[33]** functions. The fusion of these technologies with body tracking and motion compression algorithms allow a very high immersive level of virtual presence, which established VR systems like CAVEs and Panoramic Walls cannot compete with. The high immersion in combination with realistic 3D documented media artwork is the reason why our research addresses telepresence technologies and also prefers this terminology instead of the widespread VR notion of "a simulation of physical presence." Indeed, in an e-Installation there is no simulation but rather a realistic interaction with a "living document" that re-enacts all features of the real artifact. Moreover, the chance to interact with and observe other remote visitors is a quality indispensable for perceiving the "blind spot" of one's own perception. In this way, the visitor can become a second-order observer of the art system, like in real life.

The e-Installation as telepresence-based documentation builds a new category of media art documentation and conservation. As a new method, it will take time to accurately determine what kind of media art can and should be documented with it. For this purpose, many experiments with different works of art will be necessary. Even more difficult is the standardization of measures and steps that have to be performed to re-enact a work of media art, because every artwork has its own very specific features and representation claims. For now, we can say that media artwork with kinetic and audio-visual elements as well as wide-ranging art installations like land art are especially suited for conversion into virtual 3D art environments that can be visited with a convenient telepresence system. In particular, kinetic and sound artwork by artists such as Jean Tingely, Alexander Calder, Nam June Paik, Rebecca Horn, Jeffrey Shaw, Stephan von Huene, but also temporary modern art installations like Christo's wrapped buildings, Ólafur Elíasson's artificial waterfalls, or even Per Barclay's liquid installations come into consideration for e-Installation.

For every media artwork that has to be synesthetically documented, it is necessary to carry out a detailed investigation of its meaning, the artist's intention, its technical features, its construction plans, etc. This investigation has to be performed following systematic data collection methods according to modern art conservation practices **[5-9, 11, 34]**. The selection criteria depend on several aspects, which have to be determined by art experts and computer scientists. For the present research, we defined the following key aspects: a) the artwork's relevance in terms of art history; b) the artwork's level of vulnerability and accessibility; c) access to a documentation of the artwork with detailed information about, e.g., the artist's intention and the materials used, as well as the artwork's technological basis and its construction plans; d) the technical viability of the documentation method; and e) the conceptual and material suitability for the telepresence-based documentation method.

As in the conservation and restoration of modern and contemporary art, a "decision-making model" is needed to deduce the conservation options, i.e., the "virtualization and re-enactment options" in the case of an e-Installation. The "virtualization and re-enactment options" can be very different depending on the object and the artist's intention. Some artists attach a lot of significance to apparently trivial things while other aspects of an artwork are of much less importance to them. Knowing these details helps avoid mistakes and misunderstandings. The transmutation of the material conditions of an artwork through its virtualization can also change its meaning, so the relation of the physical conditions of an artwork to its meaning must be investigated before a virtualization treatment is proposed. If there is no connection between the material conditions of an artwork and its meaning, then it is possible to reproduce the basic structure of the artwork without laborious photogrammetric methods or the use of sensor data for textures. On the other hand, if the material conditions of an artwork are essential for its meaning, then it is necessary to reproduce it with high accuracy using textures, photogrammetric techniques, and so on.

## 5. The Telepresent System and the Case Studies "Versailles Fountain" by Nam June Paik and "10,000 Moving Cities – Same but Different" by Marc Lee

In order to explain, how the virtualization and the visualization in a telepresence system work, two case studies were carried out. For the present research, we used the proprioceptive extended-range telepresence system **[35]** (Fig. 1) of the Intelligent Sensor-Actuator Laboratory (ISAS) at the Karlsruhe Institute of Technology (KIT).

The telepresence system at ISAS offers a broad and very adequate experimental ground for testing and developing realistic art scenarios. It consists of three basic components: 1) a server PC that runs the VR-application and the telepresence framework; 2) an HMD (Head Mounted Display) connected to a wireless client PC, worn by the user, which streams the data to be displayed from the server; and 3) a tracking system that allows the user to navigate and interact within the telepresence system. Here are more details about the three basic components:

1) The server PC:
The server PC consists of Open GL Rendering, a Blender-compatible 3D model, Python, and a C++ interface for program logic.

2) The HMD connected to the client PC:
The HMD is an Oculus Rift with surround-sound headphones, and is connected to a backpack PC that streams the audio-visual information from the server PC.

3) The tracking system:
Eight Microsoft Kinect Devices capture the user's head pose, i.e., position and orientation, in an area of five square meters. The user can walk freely within this area, and the multimodal information from the VR is displayed.

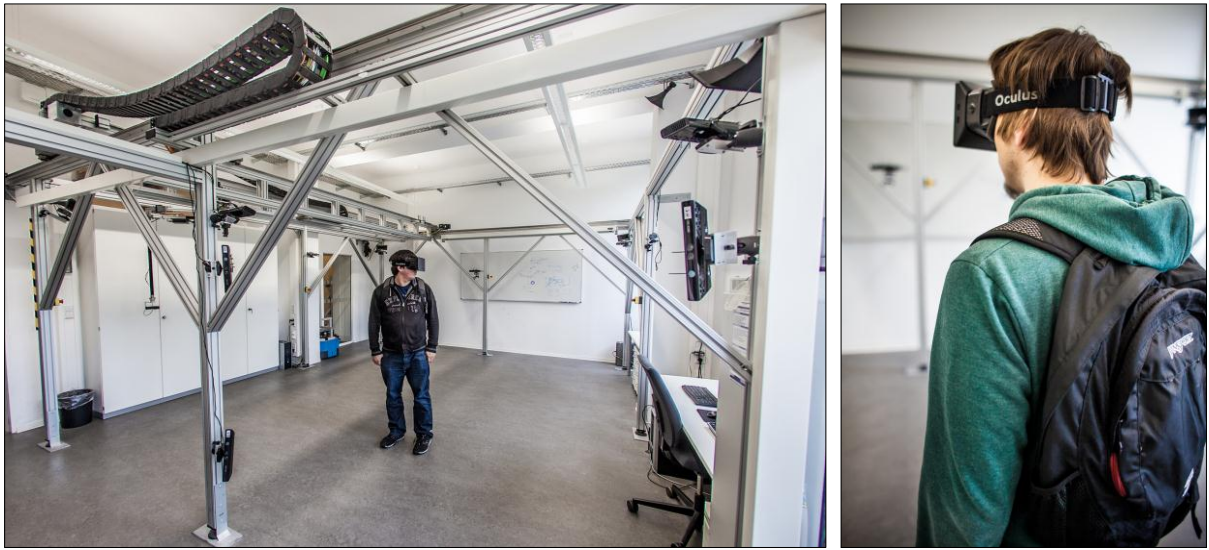

(a)  (b)

*Fig. 1:* (a) *Telepresence System at ISAS.*
(b) *Visualization System: HMD and portable PC.*

In order to allow the experience of extended-range environments, a technique called "motion compression" **[30]** predicts the desired walking path and adjusts the visual input of the HMD to guide the user on a slightly transformed path that fits into the telepresence system. As an example, if the predicted path is a straight line in the virtual environment, the algorithm would lead the user on a circular path within the telepresence system.

"Versailles Fountain" (1993) by Nam June Paik is a two-channel video sculpture that can be visited at the ZKM Karlsruhe. The lavish fountains of Versailles are used here as a metaphor for the entertainment system of our time. It consists of 20 neon and 38 CRT monitors in various sizes. The TV monitors are switched in two different circuits producing a barely perceptible half-second time-delay between UHF (ultrahigh frequency) and composite video connections.

The following criteria were taken into account in selecting this specific artwork as a case study:
1) Endangered artwork: the sculpture consists of old neon lamps and CTR televisions, and is not always accessible to the public due to preservation work being done on it.
2) Relevance: it is a relevant work by Nam June Paik, acclaimed as one of the "fathers of video art."

3) Accessibility: there is little audio-visual information (mostly just photos) about this artwork, and the work is not always available to art experts and the interested public. A temporary absence of this artwork due to maintenance or lending the artwork to other exhibitions could be bridged with the performance of its digitized version.
4) Indirect conservation: the availability of the artwork as a digital surrogate could increase the lifespan of the original.

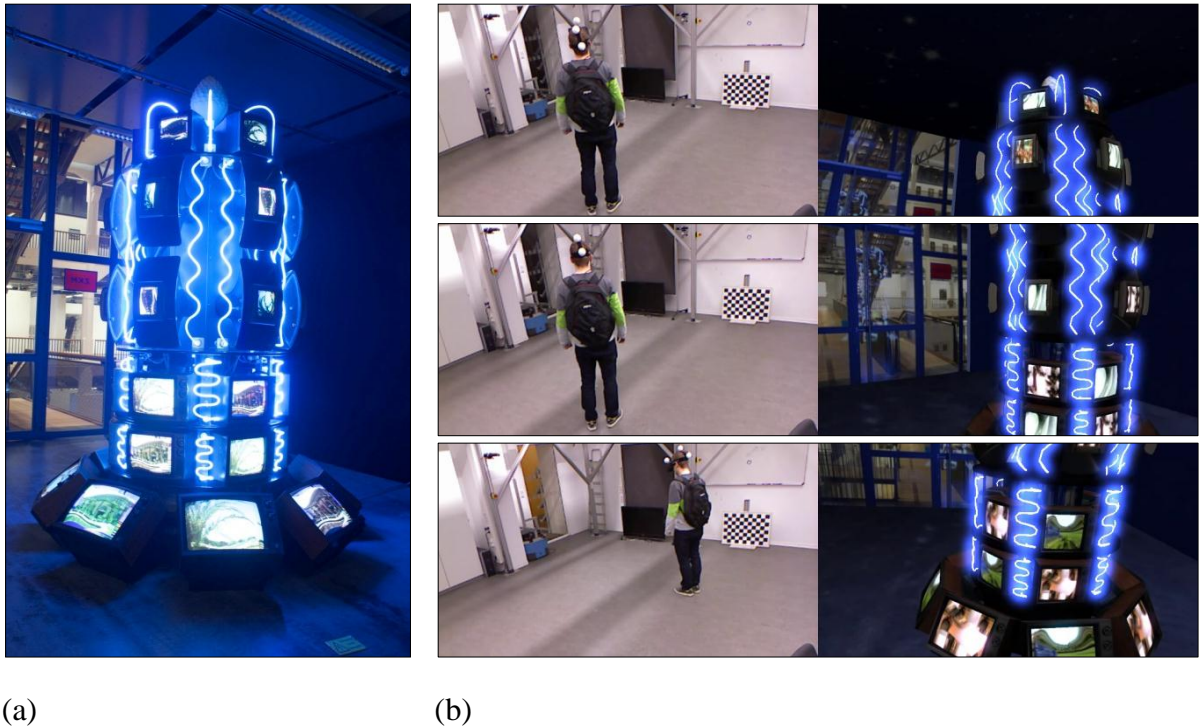

(a) (b)

*Fig. 2:* (a) *Video sculpture "Versailles Fountain" at the ZKM.*
(b) *Virtualized version in the telepresence system.*

Concerning "10,000 Moving Cities – Same but Different" (2012) by Marc Lee, the artist's intention to fully virtualize the artwork was the starting point for the case study. Furthermore, net art is a jeopardized form of cultural heritage because of its dependence on the World Wide Web, which brought about additional challenges for the case study, such as the creation of an encapsulated offline version.

For the identification of authenticity criteria and a decision-making model in Marc Lee's "10,000 Moving Cities – Same but Different," a questionnaire by Fabienne Blanc about a similar work **[36]** was enhanced for a detailed interview with the artist that focused on conservation issues and exhibit requirements, among other things.

With reference to "Versailles Fountain," the research on the art piece was complemented by the specific knowledge of the relevant curator and technical personal at the ZKM museum. Once information relating to art history, material conditions, technical documentation, and the artist's intention were carefully examined and contrasted, a treatment model for 3D modeling and telepresent re-enactment was designed with a focus on computational complexity reduction, while still respecting the requirements and authenticity criteria for digital re-enactment.

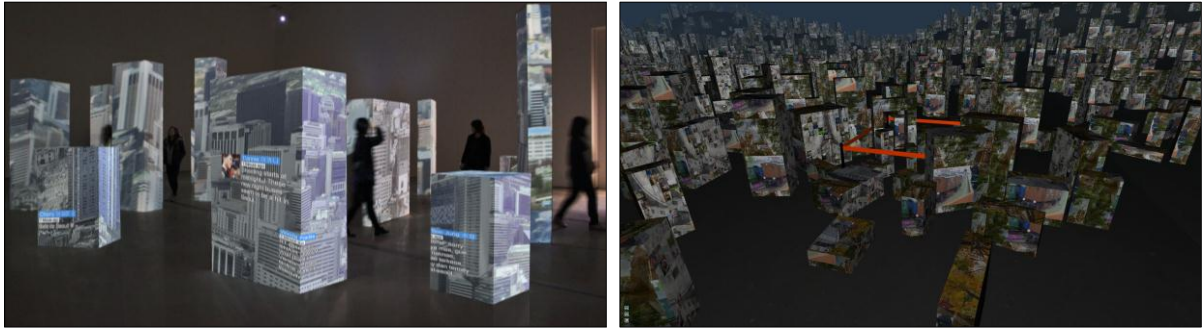

(a)                                                    (b)

*Fig. 3:* (a) *Net artwork "10,000 Moving Cities – Same but Different" at Seoul.*
        (b) *Virtualized version in the telepresence system.*

According to the experiences with "Versailles Fountain" (VF) and "10,000 Moving Cities – Same but Different" (MC), the technical process of generating an e-Installation can be structured into four essential parts:
1) creating a geometric 3D model,
2) using sensor data to model the more complex parts,
3) incorporating existing digital content, and
4) developing program logic.

*1) 3D Modeling*
The simpler parts, such as the cubes (MC) and the body (VF) of the original installations were carefully modeled in Blender. For the "Versailles Fountain," the body of the artwork was found to be simple enough to apply this type of modeling.

*2) Sensor Data*
However, the more complex parts, as well as the textures, were reconstructed from real data captured from the installation. In particular, we applied Structure From Motion according to **[37]** in order to reconstruct the 3D geometry of the pinecone on top of the artwork (see Figure 2). In addition, a panoramic image of the surroundings was created in order to embed the VF into its exhibition environment at the ZKM.

*3) Digital Content*
Existing digital content, i.e., video files (VF) and code (MC), was adapted and incorporated.
In VF, two video files, stored on the hard disk of a PC, are played in a loop on the television sets. MC contains five browser applications written in HTML, where four of them generate audio-visual collages of current online content that are then projected onto the cubes. Another application, which implements a menu based on Google Maps, also requires a connection to the Internet.

*4) Developing Program Logic*
We implemented the program logic for both e-Installations in Python. For the VF, the logic simply required displaying the video files in the form of video textures on the virtual televisions in the 3D model of the fountain. In the case of MC, implementing the logic was much more complex as it required (1) displaying current browser windows as a video texture and (2) allowing for interaction with the menu. This included implementing the menu for selecting cities, as well as implementing virtual speakers and virtual projectors that project video textures onto the cubes. This has been challenging from a technical perspective, as all the digital content of MC is loaded online from the Internet, i.e., from Google Maps, Twitter, YouTube, and other sites.

Once an e-Installation is created according to the scheme described above, it can be experienced using the telepresence system at the ISAS lab. As mentioned at the beginning of this section, in the telepresence system the user is equipped with a head-mounted display as well as headphones and can freely explore the e-Installation by walking and looking around. A server PC runs the application and synthesizes the sensory impressions of the e-Installation according to the current user location and perspective. These impressions are then transmitted to the client PC and rendered to the user.

The synesthetic documentation of "Versailles Fountain" allows unrestricted access to a very realistic model of the sculpture and its meaning level for everyone everywhere. In this scenario, the virtualization and telepresent visualization of a static video sculpture with embedded original video signals was tested.

"10,000 Moving Cities" provided a different scenario with real-time data from the Internet and interaction with a search interface. The cooperation with the Swiss net artist Marc Lee has shown that an e-Installation also offers a very attractive alternative for exhibiting immersive net art beyond the conservation of synesthetic documentation.

## 6. Challenges of Digital "Re-enactment" beyond Technical Issues: Curatorial Decisions and the e-Installation Paradox

During the digital re-enactment of a work of media art and its transfer into a telepresence system, some curatorial decisions must be taken. Curatorial input led to the implementation of the environment as well as to relevant decisions concerning detail modeling.

*Environment*
In both cases, an environment was recreated in order to preserve contextual information: the real exhibition place in the case of VF, and an imaginary landscape conceived by the artist in the case of MC. This recreation helps document the intention of the curator and the artist, respectively.

*Modeling Detail*
Ideally, the entire composition of the sculpture, including hardware, circuit, and wiring diagrams, has to be captured. The following questions regarding the level of modeling detail have therefore been addressed:
1) What level of detail has to be achieved in the modeling of hardware components?
2) Should unintended side effects such as a half-second time delay between the UHF and the composite video signals of VF be emulated?

After discussions with the curator and the ZKM technicians, it was clear that the position of the knobs of VF has no effect on the images. This means that the modeling of the rotation of the knobs can be omitted without the sculpture losing authenticity at the level of meaning. In the case of the half-second time delay between the monitors in VF, this was not originally intended by Nam June Paik and is only perceptible on closer inspection, so there was no need to emulate this.

*The e-Installation Paradox*

The creation of such a complex digital artifact, i.e., an e-Installation, is an interesting paradox for the conservation of media art: the synesthetic documentation becomes an artwork itself, and needs, for its part, a preventive preservation framework in order to keep functioning when the software context and the hardware configuration change. In particular, the virtualization of physical components of digital-born artwork, as in net art or game art, does not guarantee a long-term solution for the conservation of meanings and processes. For this purpose, continuous maintenance with regular updates and adaptation of all involved software components is required. Therefore, e-Installations are also the objects of long-term archiving experts.

## 7. Conclusions

In this paper, a new synesthetic documentation method for the virtual re-enactment of media artwork was presented and tested. This new method, which is called e-Installation, can be integrated into modern art conservation practices as a form of extended documentation within the framework of an informational preservation strategy [3]. In addition, it offers scalable access not only for curators, artists, conservators and art theorists, but also for art communicators and the general public. An e-Installation also provides 3D modeling and telepresence experts a large field of research on human perception thresholds, which influence the complexity and resolution of the virtual re-enactment.

We have seen that a realistic and useful e-Installation implies a deep knowledge about the artwork and its authenticity criteria. This is only possible if systematic modern art documentation and conservation planning methods such as the "variable media questionnaire" and the "decision-making model" (section 4) are first taken into account. In this way, mistakes can certainly be avoided during the virtual re-enactment.

Two scenarios have so far been tested: the video sculpture "Versailles Fountain" by Nam June Paik and the net art installation "10,000 Moving Cities – Same but Different" by Marc Lee. In both cases, four steps were followed for the technical implementation: 1) creation of 3D objects in Blender; 2) use of sensor data; 3) integration of existing digital content; and 4) the implementation of the art program logic. The resulting application runs on a server PC that synthesizes the sensory impressions of the artwork according to the current user location and perspective in a telepresence system.

The synesthetic documentation of media artwork at risk is still in its infancy. New scenarios are required to achieve new goals and define systematic decision-making models. As an example, kinetic art and land art are particularly suitable for the improvement of realistic proxemics, multiple-visitors interaction, and second-order observation. These art scenarios require further development of telepresence techniques like "haptics" and "motion compression."

The present research has also shown that an e-Installation can transcend current conservation thinking by creating an entirely new media artwork in collaboration with an artist. Finally, the use of an e-Installation within and beyond a conservation context has implications for the authenticity concept, which requires further study.

**Acknowledgments**
We would like to acknowledge the support of ZKM director Peter Weibel, curator Bernhard Serexhe, and technician Mirco Frass for facilitating access to Nam June Paik's video sculpture and for sharing their knowledge. We would also like to thank the students Jennifer McClelland, Pascal Becker, and Xuefei Zheng for their work on the re-enactment of "Versailles Fountain," as well as the Swiss net artist Marc Lee and the students Filip Szeliga, Michael Schröder, and Jan Philipp Gerlach for their cooperation in implementing "10,000 Moving Cities – Same but Different." For helpful advice, we also thank Jens Görisch and Antonia Pérez Arias.